\documentclass{llncs}
\usepackage{makeidx}  
\usepackage{amssymb}
\usepackage{amsmath}
\usepackage{url}

\usepackage{lastpage}

\usepackage{fancyhdr} 
\renewcommand{\chaptermark}[1]{\markboth{\textbf{\thechapter}\ \emph{#1}}{}}

\makeatletter
\renewcommand{\@evenfoot}%
{\hfil{\thepage} of \pageref{LastPage}\hfil}
\renewcommand{\@oddfoot}{\@evenfoot}
\makeatother

\begin{document}
\mainmatter              
\title{Fibonacci connection between Huffman codes and Wythoff array}
\author{Alex Vinokur}
%
%
\tocauthor{Holon, Israel}
\institute{Holon, Israel\\
\email{alexvn@barak-online.net}\\
\email{alex.vinokur@gmail.com}\\ Home Page:
\texttt{http://alexvn.freeservers.com/}}

\maketitle              
\hfil{Revised October 07, 2005\hfil

\begin{abstract}
A non-decreasing sequence of positive integer weights $P =\{p_{1}, \ldots, p_{2}, p_{n}\}$ 
is called $k$-$ordered$ if an intermediate sequence of weights produced 
by Huffman algorithm for initial sequence $P$ on $i$-th step satisfies 
the following conditions: $p^{(i)}_{2}$ = $p^{(i)}_{3}$, $i=\overline{0,k}$;
$p^{(i)}_{2} < p^{(i)}_{3}$, $i=\overline{k+1,n-3}$.
Let $T$ be a binary tree of size $n$ and $M=M(T)$ be a set of such sequences 
of positive $integer$ weights that the tree $T$ is the Huffman tree 
of $P$ $(|P|=n)$. A sequence $P_{\mathrm{min}}$
 of $n$ positive integer weights is called 
a $minimizing$ sequence of the binary tree $T$ in class $M (P_{\mathrm{min}} \in M)$ 
if $P_{\mathrm{min}}$ produces the minimal Huffman cost of the tree $T$ over 
all sequences from $M$, i.e., $E(T,P_{\mathrm{min}}) \le E(T,P) \ \forall P \in M$. 
Fibonacci related connection between minimizing $k$-ordered sequences of 
the maximum height Huffman tree and the Wythoff array [Sloane, A035513] has been proved.
Let $M_{n,k}$ $(k=\overline{0,n-3})$ denote the set of all $k$-ordered sequences of size $n$ 
for which the Huffman tree has maximum height. 
Let $F(i)$ denote $i$-th Fibonacci number. 
\underline{\bfseries{Theorem}}: A minimizing $k$-ordered sequence of the maximum height Huffman tree 
in class $M_{n,k}$ $(k=\overline{0,n-3})$ is 
$P_{\mathrm{min}_{n,k}} =\{p_{1}, p_{2}, \ldots, p_{n}\}$, where 
$p_{1}=1,\ p_{2} = F(1), \ldots, \ p_{k+2} = F(k+1), \ p_{k+3} = F(k+2) = w_{F(k+2),0}, 
\ p_{k+4} = w_{F(k+2),1}, \ p_{k+5} = w_{F(k+2),2}, \ldots, \ p_{n} = w_{F(k+2),n-k-3}$; 
\ $w_{i,j}$ is $(i,j)$-th element of the Wythoff array.$\blacksquare$
The cost of Huffman trees for those sequences has been computed. 
Several examples of minimizing ordered sequences for Huffman codes are shown.
\end{abstract}
\section{Main Conceptions and Terminology}
\subsection{Binary Trees}
A (strictly) $\underline{binary \ tree}$ is an oriented ordered tree where each nonleaf node 
has exactly two children (siblings). A binary tree is called $\underline{elongated}$ 
if at least one of any two sibling nodes is a leaf. 
An elongated binary tree of size $n$ has maximum height among all binary trees 
of size $n$. An elongated binary tree is called $\underline{left-sided}$ if the right node 
in each pair of sibling nodes is a leaf.

A binary tree is called $\underline{labeled}$ if a certain positive integer (weight) 
is set in correspondence with each leaf.

$\underline{\mbox{Size of a tree}}$ is the total number of $leaves$ of this tree.

\

\noindent
{\bfseries {Definition.}} \emph{Let $T$ be a binary tree with positive weights 
$P =\{p_{1}, p_{2}, \ldots, p_{n}\}$ at its leaf nodes. 
The weighted external path length of $T$ is
\begin{eqnarray*}
  E(T,P) = \sum^{n}_{i=1} l_i \cdot p_i,
\end{eqnarray*}
where $l_i$ is the length of the path from the root to leaf $i$.}

\subsection{Huffman Algorithm}

{\bfseries {Problem definition.}} Given a sequence of $n$ positive weights $P =\{p_{1}, \ldots, p_{n}\}$. 
The problem is to find binary tree $T_{\mathrm{min}}$ with $n$ leaves labeled $p_{1}, p_{2}, \ldots, p_{n}$ 
that has minimum weighted external path length over all possible binary 
trees of size $n$ with the same sequence of leaf weights. 
$T_{\mathrm{min}}$ is called the Huffman tree of the sequence $P$; $E(T,P_{\mathrm{min}})$ is called 
the Huffman cost of the tree $T$.

The problem was solved by Huffman algorithm \cite{huffman:1}. That algorithm builds 
$T_{\mathrm{min}}$ in which each leaf (weight) is associated with a (prefix free) codeword 
in alphabet $\{0, 1\}$.

$Note$. A code is called a prefix (free) code if no codeword is a prefix of another one.

\

\noindent
{\bfseries {Algorithm description}} (in the reference to the discussed issue).

\underline{Algorithm input}. A non-decreasing sequence of positive weights 
\begin {center}
 {$P =\{p_{1}, p_{2}, \ldots, p_{n}\} \ (p_k \le p_{k+1}; k=\overline{1,n-1}).$}
\end {center}

\underline{Algorithm output}. The sum of all the weights.

The algorithm is performed in $n-1$ steps. $i$-th step $(i=\overline{1,n-1})$ is as follows:
\begin{itemize}
\item {\underline{$i$-th step} input}. A non-decreasing sequence of weights of size $n-i+1$.
\begin {center}
 {$P^{(i-1)} =\{p^{(i-1)}_{1}, p^{(i-1)}_{2}, \ldots, p^{(i-1)}_{n-i+1}\} \ (p^{(i-1)}_k \le p^{(i-1)}_{k+1}; k=\overline{1,n-i}).$}
\end {center}
\item {\underline{$i$-th step} method}. Build a sequence 
 $\{p^{(i-1)}_{1}+p^{(i-1)}_{2}, p^{(i-1)}_{3},  \ldots, p^{(i-1)}_{n-i+1}\}$
 and sort its.
\item {\underline{$i$-th step} output}. A non-decreasing sequence of weights of size $n-i$.
\begin {center}
 {$P^{(i)} =\{p^{(i)}_{1}, p^{(i)}_{2}, \ldots,, p^{(i)}_{n-i}\} \ 
 (p^{(i)}_k \le p^{(i)}_{k+1}; k=\overline{1,n-i-1}).$}
\end {center}
\end{itemize}

\emph{Note 1}. $P^{(0)}$ is an input of Huffman algorithm, i.e., 
\begin{eqnarray}
  p^{(0)}_k = p_k \ (k=\overline{1,n}). \label{eq:a01}
\end{eqnarray}

\emph{Note 2}. If an input sequence on $i$-th step(s) of the algorithm satisfies condition 
\begin{eqnarray*}
  p^{(i)}_2 = p^{(i)}_3 \ (0 \le i \le n-3). \label{eq:a02}
\end{eqnarray*}
then several Huffman trees can result from initial sequence $P$ of weights, 
but the weighted external path length is the same in all these trees.

Let $P =\{p_{1}, p_{2}, \ldots, p_{n}\}$ be a sequence of size $n$ for which 
the binary Huffman tree is elongated. Then according to Huffman algorithm
\begin{eqnarray}
  p^{(i)}_1 + p^{(i)}_2 \le p^{(i)}_4 \ (i=\overline{0,n-3}).
\end{eqnarray}

\subsection{Wythoff Array}
The Wythoff array is shown below.
It has many interesting properties \cite{sloane:1}, \cite{sloane:2}, \cite{fra:kimb}.

\begin{center}
\begin{tabular}{||c|c||r||r|r|r|r|r|r|r|r|r|r|r||l||}
\hline
\multicolumn{2}{||c||}{}&1&2&3&4&5&6&7&8&9&10&11&12&\multicolumn{1}{|c||}{}\\
\cline{4-14}
\multicolumn{2}{||c||}{Row number}&\multicolumn{1}{c||}{}&\multicolumn{11}{c||}{Wythoff array}&\multicolumn{1}{c||}{Note}\\

\hline\hline
&0&1&1&2&3&5&8&13&21&34&55&89&144&Fibonacci seq\\
Fib[2]&1&3&4&7&11&18&29&47&76&123&199&322&521&Lucas seq\\
Fib[3]&2&4&6&10&16&26&42&68&110&178&288&466&754&\\
Fib[4]&3&6&9&15&24&39&63&102&165&267&432&699&1131&\\
&4&8&12&20&32&52&84&136&220&356&576&932&1508&\\
Fib[5]&5&9&14&23&37&60&97&157&254&411&665&1076&1741&\\
&6&11&17&28&45&73&118&191&309&500&809&1309&2118&\\
&7&12&19&31&50&81&131&212&343&555&898&1453&2351&\\
Fib[6]&8&14&22&36&58&94&152&246&398&644&1042&1686&2728&\\
&9&16&25&41&66&107&173&280&453&733&1186&1919&3105&\\
&10&17&27&44&71&115&186&301&487&788&1275&2063&3338&\\
&11&19&30&49&79&128&207&335&542&877&1419&2296&3715&\\
&12&21&33&54&87&141&228&369&597&966&1563&2529&4092&\\
Fib[7]&13&22&35&57&92&149&241&390&631&1021&1652&2673&4325&\\
\hline
\multicolumn{1}{||c|}{}&\multicolumn{13}{c||}{Generalized Wythoff array}&\multicolumn{1}{c||}{}\\

\hline
\end{tabular}
\end{center}

The two columns to the left of the Wythoff array consist respectively 
of the nonnegative integers $n$, and the lower Wythoff sequence whose 
$n$-th term is $[(n+1)*\varphi]$, where $\varphi = (1+\sqrt 5)/2$ (Golden Ratio). 
The rows are then filled in by the Fibonacci rule that each term 
is the sum of the two previous terms. The entry $n$ in the first column 
is the index of that row.

$Note$. The Wythoff array description above has been taken from \cite{sloane:1}.
Let $w_{i,j}$ denote an $(i,j)$-th element of 
the genaralized Wythoff array (row number $i \ge 0$,  column number $j \ge 0$).

\subsection{Fibonacci Numbers and Auxiliary Relations}
Let $F(i)$ denote $i$-th Fibonacci number, i.e., $F(0) = 0, F(1) = 1, F(i) = F(i-1) + F(i-2)$ 
when $i > 1$, $L(i)$ denote $i$-th Lucas number, i.e. $L(1) = 1, L(2) = 3, L(i) = L(i-1) + L(i-1)$ 
when $i > 2$.
Note some property of the Wythoff array that is related to the discussed issue:
\begin{eqnarray}
  w_{F(i),j} = F(i+j)+F(j), \ i \ge 2, j \ge 0. \label{eq:a03}
\end{eqnarray}

Note also the following property of Fibonacci numbers 
\begin{eqnarray}
  1 + \sum^{i}_{j=1} F(i) = F(i+2). \label{eq:a04}
\end{eqnarray}

\section{Main Results}
Let $T$ be a binary tree of size $n$ and $M=M(T)$ be a set of such sequences 
of positive integer weights that the tree $T$ is the Huffman tree of $P (|P|=n)$.

\

\noindent
{\bfseries {Definition.}} \emph{A sequence $Pmin$ of $n$ positive integer weights 
is called a \underline{minimizing} sequence of the binary tree $T$ in class 
$M (P_{\mathrm{min}} \in M)$ if $P_{\mathrm{min}}$ produces the minimal 
Huffman cost of the tree $T$ over all sequences from $M$, i.e.,}
\begin{eqnarray*}
  E(T,P_{\mathrm{min}}) \le E(T,P) \ \forall P \in M.
\end{eqnarray*}

\noindent
{\bfseries {Definition.}} \emph{A non-decreasing sequence of positive integer weights 
\\
$P = \{p_{1}, p_{2}, \ldots, p_{n}\}$ is called \underline{absolutely ordered} if the intermediate 
sequences of weights produced by Huffman algorithm for initial 
sequence $P$ satisfy the following conditions}
\begin{eqnarray*}
 p^{(i)}_2 < p^{(i)}_3, \ i = \overline{0,n-3}.
\end{eqnarray*}

\begin{theorem} [\cite{vinokur:1986}] \label{theor:01}
A minimizing absolutely ordered sequence of the elongated binary tree is 
\begin{eqnarray*}
  P_{\mathrm{min_{abs}}} = \{F(1), F(2), ., F(n)\},
\end{eqnarray*}
where F(i) is i-th Fibonacci number.
The weighted external path length of elongated binary tree T 
of size n for the minimizing absolutely ordered sequence Pminabs is
\begin{eqnarray*}
  E(T, P_{\mathrm{min_{abs}}}) = F(n+4) - (n + 4).
\end{eqnarray*}
\end{theorem}

\begin{proof}
The proof of Theorem 1 of \cite{vinokur:1986}.\qed
\end{proof}
\noindent
{\bfseries {Definition.}} 
\emph{A non-decreasing sequence of positive integer weights \\
$P^{(i)} =\{p^{(i)}_{1}, p^{(i)}_{2},  \ldots, p^{(i)}_{n-i}\}$
is called \underline{k-ordered} if the intermediate sequences 
of weights produced by Huffman algorithm for initial sequence P 
satisfy the following conditions}
\begin{gather}
  p^{(i)}_2 = p^{(i)}_3 \ i=\overline{0,k}; \label{eq:a05}\\
  p^{(i)}_2 < p^{(i)}_3 \ i=\overline{k+1,n-3}.	\label{eq:a06}
\end{gather}

Let $M_{n,k} (k=\overline{0,n-3})$ denote the set of all $k$-ordered sequences 
of size $n$ for which the binary Huffman tree is elongated, 
i.e. an elongated binary tree of size $n$ is the Huffman tree of $P$.

\begin{theorem}	\label{theor:02}
A minimizing k-ordered sequence of the \underline{elongated} binary tree 
in class $M_{n,k} (k=\overline{0,n-3})$ is 
\begin{gather*}
p_1 = 1,\\
p_{i} = F(i-1), \ i = \overline{2,k+2},\\
p_{k+3} = F(k+2) = w_{F(k+2),0}, \\
p_{i} = w_{F(k+2),i-k-3}, \ i = \overline{k+4,n},
\end{gather*}
where $w_{i,j}$ is $(i,j)$-th element of the Wythoff array.
\end{theorem}

\begin{proof}
Because $P_{\mathrm{min}_{n,k}} =\{p_{1}, p_{2}, \ldots, p_{n}\}$
is $minimizing$ sequence of positive integer values,
$p_1$ and $p_2$ should have minimal positive integer values, i.e.,
\begin{eqnarray}
p_1 = p_2 = 1.\label{eq:a07}
\end{eqnarray}
$P_{\mathrm{min}_{n,k}}$ is $k$-ordered $(k \ge 0 )$ sequence, so according to (\ref{eq:a05})
\begin{eqnarray*}
p_2 = p_3 = 1,
\end{eqnarray*}
therefore according to (\ref{eq:a01}) and (\ref{eq:a07})
\begin{eqnarray*}
p_3 = 1.
\end{eqnarray*}
\ \ \ \ \ Thus
\begin{eqnarray}
p_1 = 1, \ p_2 = F(1), \ p_3 = F(2). \label{eq:a08}
\end{eqnarray}
Further, taking into account (\ref{eq:a02}) and (\ref{eq:a06}) 
we obtain the following Huffman algorithm steps for $k$-ordered $(k \ge 0)$ 
sequence of the elongated (left-sided) binary tree.

\begin{center}

\begin{tabular}{c | l | l}
\hline
\hline
\multicolumn{3}{c}{Steps $0-k$}\\
\hline
\hline
Step & Input sequence & Relation \\
\hline
0 & $p_1, p_2, p_3, p_4, p_5, p_6,    \ldots, p_{k-1}, p_k, p_{k+1}, \ldots, p_{n-1}, p_n$; & $p_2 = p_3$ \\
1 & $p_3, \sum\limits^{2}_{i=1} p_i, p_4, \ldots, p_{k-1}, p_k, p_{k+1}, \ldots, p_{n-1}, p_n$; & $\sum\limits^{2}_{i=1} p_i = p_4$ \\
2 & $p_4, \sum\limits^{3}_{i=1} p_i, p_5, \ldots, p_{k-1}, p_k, p_{k+1}, \ldots, p_{n-1}, p_n$; & $\sum\limits^{3}_{i=1} p_i = p_5$ \\
3 & $p_5, \sum\limits^{4}_{i=1} p_i, p_6, \ldots, p_{k-1}, p_k, p_{k+1}, \ldots, p_{n-1}, p_n$; & $\sum\limits^{4}_{i=1} p_i = p_6$ \\
\dots & & \\
$k$-1 & $p_{k+1}, \sum\limits^{k}_{i=1} p_i, p_{k+2}, \ldots, p_{n-1}, p_n$; & $\sum\limits^{k}_{i=1} p_i = p_{k+2}$ \\
$k$ & $p_{k+2}, \sum\limits^{k+1}_{i=1} p_i, p_{k+3}, \ldots, p_{n-1}, p_n$; & $\sum\limits^{k+1}_{i=1} p_i = p_{k+3}$ \\
\end{tabular}

\

\

\

\begin{tabular}{c | l | l}
\hline
\hline
\multicolumn{3}{c}{Steps $(k+1)-(n-3)$}\\
\hline
\hline
Step & Input sequence & Relation \\
\hline
$k$+1 & $p_{k+3}, \sum\limits^{k+2}_{i=1} p_i, p_{k+4}, p_{k+3}, \ldots, p_{n-1}, p_n$; & $\sum\limits^{k+2}_{i=1} p_i < p_{k+4}$ \\
$k$+2 & $p_{k+4}, \sum\limits^{k+3}_{i=1} p_i, p_{k+5}, \ldots, p_{n-1}, p_n$; & $\sum\limits^{k+3}_{i=1} p_i < p_{k+5}$ \\
\dots & & \\
$n$-4 & $p_{n-2}, \sum\limits^{n-3}_{i=1} p_i, p_{n-1}, p_n$; & $\sum\limits^{n-3}_{i=1} p_i < p_{n-1}$ \\
$n$-3 & $p_{n-1}, \sum\limits^{n-2}_{i=1} p_i, p_n$; & $\sum\limits^{n-2}_{i=1} p_i < p_{n}$ \\
\end{tabular}

\

\

\

\begin{tabular}{c | l}
\hline
\hline
\multicolumn{2}{c}{Steps $(n-2),(n-1)$}\\
\hline
\hline
Step & Input sequence \\
\hline
$n$-2 & $p_n, \sum\limits^{n-1}_{i=1} p_i$\\
$n$-1 & $\sum\limits^{n}_{i=1} p_i$\\
\end{tabular}

\end{center}

Consider two cases.

{\bfseries{\underline{Case 1}. Steps $0-k$.}}

It follows from relations for steps $0-k$ that
\begin{eqnarray*}
  p_i = \sum^{i-2}_{j=1} p_j, \ i=\overline{4,k+3}.
\end{eqnarray*}
Thus,
\begin{eqnarray*}
  p_i - p_{i-1}= \sum^{i-2}_{j=1} p_j - \sum^{i-3}_{j=1} p_j = p_{i-2}, \ i=\overline{4,k+3}.
\end{eqnarray*}
So, we have 
\begin{eqnarray*}
  p_i = p_{i-1} + p_{i-2}, \ i=\overline{4,k+3}.
\end{eqnarray*}
Taking into account (\ref{eq:a08}), we obtain
\begin{eqnarray}
p_i = F(i-1). \label{eq:a09}
\end{eqnarray}

In particular,
\begin{eqnarray}
p_{k+3} = F(k+2) = F(k+2) + F(0). \label{eq:a10}
\end{eqnarray}

{\bfseries{\underline{Case 2}. Steps $(k+1)-(n-3)$.}}

Because $P_{\mathrm{min}_{n,k}} = \{p_1,p_2,\ldots,p_n\}$
is $line{minimizing}$ sequence of positive ${integer}$ values, 
inequalities for steps $(k+1)-(n-3)$ are transformed 
to the following equalities:

\

\begin{tabular}{c | l | l}
\hline
\hline
\multicolumn{3}{c}{Steps $(k+1)-(n-3)$}\\
\hline
\hline
Step & Input sequence & Relation \\
\hline
$k$+1 & $p_{k+3}, \sum\limits^{k+2}_{i=1} p_i, p_{k+4}, p_{k+3}, \ldots, p_{n-1}, p_n$; & $\sum\limits^{k+2}_{i=1} p_i = p_{k+4}+1$ \\
$k$+2 & $p_{k+4}, \sum\limits^{k+3}_{i=1} p_i, p_{k+5}, \ldots, p_{n-1}, p_n$; & $\sum\limits^{k+3}_{i=1} p_i = p_{k+5}+1$ \\
\dots & & \\
$n$-4 & $p_{n-2}, \sum\limits^{n-3}_{i=1} p_i, p_{n-1}, p_n$; & $\sum\limits^{n-3}_{i=1} p_i = p_{n-1}+1$ \\
$n$-3 & $p_{n-1}, \sum\limits^{n-2}_{i=1} p_i, p_n$; & $\sum\limits^{n-2}_{i=1} p_i = p_{n}+1$ \\
\end{tabular}

\

From the equality for step $(k+1)$, (\ref{eq:a09}), (\ref{eq:a07}) and (\ref{eq:a04}) results 
\begin{eqnarray}
p_{k+4} = F(k+3) + 1 = F(k+3) + F(1). \label{eq:a11}
\end{eqnarray}

Further, it follows from relations with equalities for steps $(k+1)-(n-3)$ that
\begin{eqnarray*}
p_i = 1 + \sum^{i-2}_{j=1} p_j, \ i = \overline{k+5,n}.
\end{eqnarray*}

Thus,
\begin{eqnarray*}
p_i - p_{i-1} = ({1 + \sum^{i-2}_{j=1} p_j}) - ({1 + \sum^{i-3}_{j=1} p_j}) = p_{i-2}, \ i = \overline{k+5,n}.
\end{eqnarray*}
So, we have
\begin{eqnarray*}
p_i = p_{i-1} + p_{i-2}.
\end{eqnarray*}
Therefore, taking into account (\ref{eq:a10}) and (\ref{eq:a11}), we have
\begin{eqnarray*}
p_i = F(i - 1) + F(i - k - 3).
\end{eqnarray*}
From this and (\ref{eq:a03}) it follows that
\begin{eqnarray}
p_i = w_{F(k+2),i-k-3}. \label{eq:a12}
\end{eqnarray}
where $w_{i,j}$ is $(i,j)$-th element of the Wythoff array.

The statement of the theorem follows from (\ref{eq:a07}), (\ref{eq:a09}) and (\ref{eq:a12}). \qed
\end{proof}

\begin{corollary}
A minimizing $0$-ordered sequence of size $n$ for the elongated binary tree 
in $M_{n,0}$ is the Lucas sequence shifted two places right, i.e. 
\begin{eqnarray*}
\{1, 1, L(1), L(2), \ldots, L(n-2)\},
\end{eqnarray*}
where $L(i)$ is $i$-th Lucas number.
\end{corollary}

\begin{corollary}
A minimizing $(n-3)$-ordered sequence of size $n$ for the elongated binary tree 
in $M_{n,n-3}$ is the Fibonacci sequence shifted one place right, i.e. 
\begin{eqnarray*}
\{1, F(1), F(2), \ldots, F(n-1)\},
\end{eqnarray*}
where $F(i)$ is $i$-th Fibonacci number.
\end{corollary}

Note that normalized $(n-3)$-ordered sequence of size $n$
\begin{eqnarray*}
\{1/F(n+1), F(1)/F(n+1), F(2)/F(n+1), \ldots, F(n-1)/F(n+1)\}
\end{eqnarray*}
has maximum weighted external path length over all possible 
normalized sequences of size $n$ for which Huffman tree is elongated \cite{vinokur:1993}.

\begin{theorem} \label{theor:03}
The weighted external path length of the elongated binary tree $T$ 
of size $n$ for the minimizing $k$-ordered sequence $P_{\mathrm{min}_{n,k}}$ is
\begin{eqnarray*}
E(T, P_{\mathrm{min}_{n,k}}) = F(n + 3) + F(n - k + 1) - (n - k + 3).
\end{eqnarray*}
\end{theorem}

\begin{proof}
Let $P_{\mathrm{min}_{n,k}} = \{p_{1}, p_{2}, \ldots, p_{n}\}$
be the minimizing $k$-ordered sequence of the elongated binary tree $T$ 
of size $n$.

According to Theorem~\ref{theor:02} \ $P_{\mathrm{min}_{n,k}}$ = 
\begin{eqnarray*}
\{1, F(1), \ldots, F(k+2), F(k+3)+F(1), \ldots, F(n-1)+F(n-k-3)\}.
\end{eqnarray*}
\ \ \ \ \ Weighted external path length $E(T,P_{\mathrm{min}_{n,k}}$ is 
\begin{eqnarray*}
  E(T,P_{\mathrm{min}_{n,k}}) = \sum^{n}_{i=1} l_i \cdot p_i,
\end{eqnarray*}
where $l_i$ is the length of the path from the root to leaf $i$.

$T$ is the elongated binary tree, therefore $l_1 = n - 1, l_i = n - i + 1 \ (i = \overline{2,n})$.

Then
\begin{gather*}
E(T, P_{\mathrm{min}_{n,k}}) = \sum_{i=1}^{n} l_i \cdot p_i = 
(n - 1) \cdot p_1 + \sum_{i=2}^{n} (n-i+1) \cdot p_i = \\
(n - 1) + \sum_{i=2}^{k+3} (n-i+1) \cdot F(i-1) + \\ \sum_{i=k+4}^{n} (n-i+1) \cdot (F(i-1) + F(i-k-3)) = \\
(n - 1) + \sum_{i=1}^{k+2} (n-i) \cdot F(i) + \sum_{i=k+3}^{n-1} (n-i) \cdot (F(i) + F(i-k-2)) = \\
(n - 1) + \sum_{i=1}^{n-1} (n-i) \cdot F(i) + \sum_{i=k+3}^{n-1} (n-i) \cdot F(i-k-2) = \\
(n - 1) + \sum_{i=1}^{n-1} (n-i) \cdot F(i) + \sum_{i=1}^{n-k-3} (n-k-i-2) \cdot F(i) = \\
(n - 1) + \sum_{i=1}^{n-1} \sum_{j=1}^{n-i} F(i) + \sum_{i=1}^{n-k-3} \sum_{j=1}^{n-k-i-2} F(i) = \\
(n - 1) + \sum_{j=1}^{n-1} \sum_{i=1}^{j} F(i) + \sum_{j=1}^{n-k-3} \sum_{i=1}^{j} F(i).
\end{gather*}
Thus, taking into account (\ref{eq:a04}), we obtain
\begin{gather*}
E(T, P_{\mathrm{min}_{n,k}}) = (n - 1) + \sum_{j=1}^{n-1} (F(j+2)-1) + \sum_{j=1}^{n-k-3} (F(j+2)-1) = \\
\sum_{j=3}^{n+1} F(j) + \sum_{j=3}^{n-k-1} F(j) - (n-k-3) = \\
(\sum_{j=1}^{n+1} F(j)-(F(2)-F(1)))+(\sum_{j=1}^{n-k-1} F(j)-(F(2)-F(1)))-(n-k-3) = \\
= (F(n+3)-1-F(3))+ (F(n-k+1)-1-F(3))-(n-k-3) = \\
(F(n + 3) - 3) + (F(n - k + 1) - 3) - (n - k - 3) = \\
F(n + 3) + F(n - k + 1) - (n - k + 3).
\end{gather*}
The statement of the theorem proved. \qed
\end{proof}
\begin{corollary}
The weighted external path length of the elongated binary tree $T$ 
of size $n$ for the minimizing $0$-ordered sequence $P_{\mathrm{min}_{n,0}}$ 
(the Lucas sequence shifted two places right) is
\begin{eqnarray*}
E(T, P_{\mathrm{min}_{n,0}}) = F(n + 3) + F(n + 1) - (n + 3).
\end{eqnarray*}
\end{corollary}
\begin{corollary}
The weighted external path length of the elongated binary tree $T$ 
of size $n$ for the minimizing $(n - 3)$-ordered sequence $P_{\mathrm{min}_{n,n-3}}$
(the Fibonacci sequence shifted one place right) is
\begin{gather*}
E(T, P_{\mathrm{min}_{n,n-3}}) = F(n + 3) + F(n - (n - 3) + 1) - (n - (n - 3) + 3) = \\
F(n + 3) + F(4) - 6 = F(n + 3) - 3.
\end{gather*}
\end{corollary}
\section{Examples}
Several examples of minimizing ordered sequences for Huffman codes 
are shown below. 
An underlined integer in the tables means a nonleaf node 
cost obtained as a result of merging two leaf nodes on the previous step 
of the Huffman algorithm.
\begin{example}
\begin{center}
Absolutely minimizing ordered sequence $P_{\mathrm{min}_{abs}}$ of size 10
\begin{tabular}{|c|c|r|r|r|r|r|r|r|r|r|r|}
\hline
\hline
Step & $P^{(i)}$ & $p^{(i)}_1$ & $p^{(i)}_2$ & $p^{(i)}_3$ & $p^{(i)}_4$ & $p^{(i)}_5$ & $p^{(i)}_6$ & $p^{(i)}_7$ & $p^{(i)}_8$ & $p^{(i)}_9$ & $p^{(i)}_{10}$ \\
\hline
\hline
0 & $P^{(0)}$ & 1 & 1 & 2 & 3 & 5 & 8 & 13 & 21 & 34 & 55 \\
\hline
1 & $P^{(1)}$ & 2 & \underline{2} & 3 & 5 & 8 & 13 & 21 & 34 & 55 \\
\cline {1-11}
2 & $P^{(2)}$ & 3 & \underline{4} & 5 & 8 & 13 & 21 & 34 & 55 \\
\cline {1-10}
3 & $P^{(3)}$ & 5 & \underline{7} & 8 & 13 & 21 & 34 & 55  \\
\cline {1-9}
4 & $P^{(4)}$ & 8 & \underline{12} & 13 & 21 & 34 & 55 \\
\cline {1-8}
5 & $P^{(5)}$ & 13 & \underline{20} & 21 & 34 & 55 \\
\cline {1-7}
6 & $P^{(6)}$ & 21 & \underline{33} & 34 & 55 \\
\cline {1-6}
7 & $P^{(7)}$ & 34 & \underline{54} & 55 \\
\cline {1-5}
8 & $P^{(8)}$ & 55 & \underline{88} \\
\cline {1-4}
9 & $P^{(9)}$ & 143 \\
\cline {1-3}
\end{tabular}
$\{p_1, p_2, p_3, p_4, p_5, p_6, p_7, p_8, p_9, p_{10}\}$ = \{1, 1, 2, 3, 5, 8, 13, 21, 34, 55\} - \\
Fibonacci sequence of size 10.

\ \\
\end{center}
\end{example}

\begin{example}
\begin{center}
Minimizing $0$-ordered sequence $P_{\mathrm{min}_{10,0}}$
\begin{tabular}{|c|c|r|r|r|r|r|r|r|r|r|r|}
\hline
\hline
Step & $P^{(i)}$ & $p^{(i)}_1$ & $p^{(i)}_2$ & $p^{(i)}_3$ & $p^{(i)}_4$ & $p^{(i)}_5$ & $p^{(i)}_6$ & $p^{(i)}_7$ & $p^{(i)}_8$ & $p^{(i)}_9$ & $p^{(i)}_{10}$ \\
\hline
\hline
0 & $P^{(0)}$ & 1 & 1 & 1 & 3 & 4 & 7 & 11 & 18 & 29 & 47 \\
\hline
1 & $P^{(1)}$ & 1 & \underline{2} & 3 & 4 & 7 & 11 & 18 & 29 & 47 \\
\cline {1-11}
2 & $P^{(2)}$ & 3 & \underline{3} & 4 & 7 & 11 & 18 & 29 & 47 \\
\cline {1-10}
3 & $P^{(3)}$ & 4 & \underline{6} & 7 & 11 & 18 & 29 & 47 \\
\cline {1-9}
4 & $P^{(4)}$ & 7 & \underline{10} & 11 & 18 & 29 & 47 \\
\cline {1-8}
5 & $P^{(5)}$ & 11 & \underline{17} & 18 & 29 & 47 \\
\cline {1-7}
6 & $P^{(6)}$ & 18 & \underline{28} & 29 & 47 \\
\cline {1-6}
7 & $P^{(7)}$ & 29 & \underline{46} & 47 \\
\cline {1-5}
8 & $P^{(8)}$ & 47 & \underline{75} \\
\cline {1-4}
9 & $P^{(9)}$ & 122 \\
\cline {1-3}
\end{tabular}
$\{p_2, p_3\} = \{1, 1\} = \{F(1), F(2)\}$ - Fibonacci sequence of size 2; \\
$\{p_3, p_4, p_5, p_6, p_7, p_8, p_9, p_{10}\} = \{1, 3, 4, 7, 11, 18, 29, 47\}$ - \\
Wythoff array row-1 (row-F(2)) sequence of size 8 (the Lucas sequence).

\ \\
\end{center}
\end{example}

\begin{example}
\begin{center}
Minimizing $1$-ordered sequence $P_{\mathrm{min}_{10,1}}$
\begin{tabular}{|c|c|r|r|r|r|r|r|r|r|r|r|}
\hline
\hline
Step & $P^{(i)}$ & $p^{(i)}_1$ & $p^{(i)}_2$ & $p^{(i)}_3$ & $p^{(i)}_4$ & $p^{(i)}_5$ & $p^{(i)}_6$ & $p^{(i)}_7$ & $p^{(i)}_8$ & $p^{(i)}_9$ & $p^{(i)}_{10}$ \\
\hline
\hline
0 & $P^{(0)}$ & 1 & 1 & 1 & 2 & 4 & 6 & 10 & 16 & 26 & 42 \\
\hline
1 & $P^{(1)}$ & 1 & \underline{2} & 2 & 4 & 6 & 10 & 16 & 26 & 42 \\
\cline {1-11}
2 & $P^{(2)}$ & 2 & \underline{3} & 4 & 6 & 10 & 16 & 26 & 42 \\
\cline {1-10}
3 & $P^{(3)}$ & 4 & \underline{5} & 6 & 10 & 16 & 26 & 42 \\
\cline {1-9}
4 & $P^{(4)}$ & 6 & \underline{9} & 10 & 16 & 26 & 42 \\
\cline {1-8}
5 & $P^{(5)}$ & 10 & \underline{15} & 16 & 26 & 42 \\
\cline {1-7}
6 & $P^{(6)}$ & 16 & \underline{25} & 26 & 42 \\
\cline {1-6}
7 & $P^{(7)}$ & 26 & \underline{41} & 42 \\
\cline {1-5}
8 & $P^{(8)}$ & 42 & \underline{67} \\
\cline {1-4}
9 & $P^{(9)}$ & 109 \\
\cline {1-3}
\end{tabular}
$\{p_2, p_3, p_4\} = \{1, 1, 2\} = \{F(1), F(2), F(3)\}$ - Fibonacci sequence of size 3; \\
$\{p_4, p_5, p_6, p_7, p_8, p_9, p_{10}\} = \{2, 4, 6, 10, 16, 26, 42\}$ -
Wythoff array row-2 (row-F(3)) sequence of size 7.

\ \\
\end{center}
\end{example}

\begin{example}
\begin{center}
Minimizing $4$-ordered sequence $P_{\mathrm{min}_{10,4}}$
\begin{tabular}{|c|c|r|r|r|r|r|r|r|r|r|r|}
\hline
\hline
Step & $P^{(i)}$ & $p^{(i)}_1$ & $p^{(i)}_2$ & $p^{(i)}_3$ & $p^{(i)}_4$ & $p^{(i)}_5$ & $p^{(i)}_6$ & $p^{(i)}_7$ & $p^{(i)}_8$ & $p^{(i)}_9$ & $p^{(i)}_{10}$ \\
\hline
\hline
0 & $P^{(0)}$ & 1 & 1 & 1 & 2 & 3 & 5 & 8 & 14 & 22 & 36 \\
\hline
1 & $P^{(1)}$ & 1 & \underline{2} & 2 & 3 & 5 & 8 & 14 & 22 & 36 \\
\cline {1-11}
2 & $P^{(2)}$ & 2 & \underline{3} & 3 & 5 & 8 & 14 & 22 & 36 \\
\cline {1-10}
3 & $P^{(3)}$ & 3 & \underline{5} & 5 & 8 & 14 & 22 & 36 \\
\cline {1-9}
4 & $P^{(4)}$ & 5 & \underline{8} & 8 & 14 & 22 & 36 \\
\cline {1-8}
5 & $P^{(5)}$ & 8 & \underline{13} & 14 & 22 & 36 \\
\cline {1-7}
6 & $P^{(6)}$ & 14 & \underline{21} & 22 & 36 \\
\cline {1-6}
7 & $P^{(7)}$ & 22 & \underline{35} & 36 \\
\cline {1-5}
8 & $P^{(8)}$ & 36 & \underline{57} \\
\cline {1-4}
9 & $P^{(9)}$ & 93 \\
\cline {1-3}
\end{tabular}
$\{p_2, p_3, p_4, p_5, p_6, p_7\} = \{1, 1, 2, 3, 5, 8\} = \{F(1), F(2), F(3), F(4), F(5), F(6)\}$ - \\
Fibonacci sequence of size 6;\\
$\{p_7, p_8, p_9, p_{10}\} = \{8, 14, 22, 36\}$ -
Wythoff array row-8 (row-F(6)) sequence of size 4.

\ \\
\end{center}
\end{example}

\begin{example}
\begin{center}
Minimizing $7$-ordered sequence $P_{\mathrm{min}_{10,7}}$
\begin{tabular}{|c|c|r|r|r|r|r|r|r|r|r|r|}
\hline
\hline
Step & $P^{(i)}$ & $p^{(i)}_1$ & $p^{(i)}_2$ & $p^{(i)}_3$ & $p^{(i)}_4$ & $p^{(i)}_5$ & $p^{(i)}_6$ & $p^{(i)}_7$ & $p^{(i)}_8$ & $p^{(i)}_9$ & $p^{(i)}_{10}$ \\
\hline
\hline
0 & $P^{(0)}$ & 1 & 1 & 1 & 2 & 3 & 5 & 8 & 13 & 21 & 34 \\
\hline
1 & $P^{(1)}$ & 1 & \underline{2} & 2 & 3 & 5 & 8 & 13 & 21 & 34 \\
\cline {1-11}
2 & $P^{(2)}$ & 2 & \underline{3} & 3 & 5 & 8 & 13 & 21 & 34 \\
\cline {1-10}
3 & $P^{(3)}$ & 3 & \underline{5} & 5 & 8 & 13 & 21 & 34 \\
\cline {1-9}
4 & $P^{(4)}$ & 5 & \underline{8} & 8 & 13 & 21 & 34 \\
\cline {1-8}
5 & $P^{(5)}$ & 8 & \underline{13} & 13 & 21 & 34 \\
\cline {1-7}
6 & $P^{(6)}$ & 13 & \underline{21} & 21 & 34 \\
\cline {1-6}
7 & $P^{(7)}$ & 21 & \underline{34} & 34 \\
\cline {1-5}
8 & $P^{(8)}$ & 34 & \underline{55} \\
\cline {1-4}
9 & $P^{(9)}$ & 89 \\
\cline {1-3}
\end{tabular}
$\{p_2, p_3, p_4, p_5, p_6, p_7, p_8, p_9, p_{10}\} = \{1, 1, 2, 3, 5, 8, 13, 21, 34\} = \{F(1), F(2), F(3), F(4), F(5), F(6), F(7), F(8), F(9)\}$ - \\
Fibonacci sequence of size 9; \\
$\{p_{10}\} = \{34\}$ -
Wythoff array row-34 (row-F(9)) sequence of size 1.

\ \\
\end{center}
\end{example}

%
%


\begin{thebibliography}{6}
%
\bibitem {huffman:1}
{\bfseries Huffman D.},
A method for the construction of minimum redundancy codes.
Proc. of the IRE {\bfseries 40} (1952) 1098--1101

\bibitem {sloane:1}
{\bfseries Sloane N.J.A.}, 
Classic Sequences In The On-Line Encyclopedia of Integer Sequences: The Wythoff Array and The Para-Fibonacci Sequence.
Published electronically at \url{http://www.research.att.com/~njas/sequences/classic.html}

\bibitem {sloane:2}
{\bfseries Sloane N.J.A.}, 
My Favorite Integer Sequences. 
Published electronically at \url{http://www.research.att.com/~njas/doc/sg.pdf}

\bibitem {fra:kimb}
{\bfseries Fraenkel A., Kimberling C.},
Generalized Wythoff arrays, shuffles and interspersions. 
Discrete Math. {\bfseries 126} (1994) 137--149

\bibitem {vinokur:1986}
{\bfseries Vinokur A.B.},
Huffman trees and Fibonacci numbers. 
Kibernetika Issue {\bfseries 6} (1986) 9-12 (in Russian), 
English translation in Cybernetics {\bfseries 22} Issue {\bfseries 6} (1986) 692--696;
\url{http://springerlink.metapress.com/link.asp?ID=W32X70520K8JJ617}

\bibitem {vinokur:1993}
{\bfseries Vinokur A.B.},
Huffman codes and maximizing properties of Fibonacci numbers. 
Kibernetika i Systemnyi Analiz Issue {\bfseries 3} (1992) 10-15 (in Russian), 
English translation in Cybernetics and Systems Analysis {\bfseries 28} Issue {\bfseries 3} (1993) 329--334;
\url{http://springerlink.metapress.com/link.asp?ID=NJ5073781H237182}

\end{thebibliography}
\end{document}